\documentclass[prl,final,twocolumn,showpacs,showkeys]{revtex4}
\usepackage{amsmath}
\usepackage{graphicx}
\usepackage{amsfonts}
\usepackage{amssymb}
\usepackage{subfigure}
\begin{document}

\title{Arbitrary Control of Entanglement between Two Superconducting Resonators}

\author{Frederick W. Strauch$^1$} \email[Electronic address: ]{Frederick.W.Strauch@williams.edu}
\author{Kurt Jacobs$^2$}
\author{Raymond W. Simmonds$^3$ }

\affiliation{$^1$Williams College, Williamstown, MA 01267, USA \\
$^2$Department of Physics, University of Massachusetts at Boston, 100 Morrissey Blvd., Boston, Massachusetts 02125, USA \\
$^3$National Institute of Standards and Technology, 325 Broadway, Boulder, Colorado 80305 USA}

\date{\today}

\begin{abstract}
We present a method to synthesize an arbitrary quantum state of two superconducting resonators.  This state-synthesis algorithm utilizes a coherent interaction of each resonator with a tunable artificial atom  to create entangled quantum superpositions of photon number (Fock) states in the resonators.  We theoretically analyze this approach, showing that it can efficiently synthesize NOON states, with large photon numbers, using existing technology.   
\end{abstract}
\maketitle

The development of 	quantum coherent systems for information processing has traditionally focused on quantum bits (qubits), in which information is stored in two quantum states of a system.  Over the past two decades many physical systems have been devised in which qubits can be addressed and manipulated, including atoms, ions, photons, and solid-state systems \cite{mikeandike}.  However, recent experiments have demonstrated that superconducting resonators---harmonic oscillators with a theoretically infinite ladder of states---can also be addressed and manipulated for quantum state storage and transfer \cite{Sillanpaa2007}.  These resonators have excellent coherence properties, and would provide a promising alternative approach to large-scale quantum information processing.   Future progress requires a theoretical study of how to efficiently generate entanglement in coupled networks of resonators.

Recent experiments have achieved arbitrary control of a single superconducting resonator.  In particular, Hofheinz {\it et al.} used a superconducting phase qubit to synthesize an arbitrary state of a single resonator \cite{Hofheinz2009}.  While previously Fock states (states of definite photon number $n$) with $n$ up to 20 had been generated \cite{Hofheinz2008}, here superposition states were created and analyzed using Wigner tomography \cite{Hofheinz2009} for photon states with $n \le 6$.  These states were synthesized using an algorithm developed by Law and Eberly \cite{Law96}, originally designed for atomic cavity-QED systems.  An important theoretical question is whether there exists a corresponding algorithm for the synthesis of an arbitrary quantum state of two resonators ($a$ and $b$), of the general form
\begin{equation}
|\Psi \rangle = \sum_{n_a=0}^{N_a} \sum_{n_b=0}^{N_b} c_{n_a, n_b} |n_a\rangle \otimes |n_b\rangle.
\end{equation}
Among these states are the maximally entangled $N$-photon states
\begin{equation}
|\Psi\rangle = \frac{1}{\sqrt{N+1}} \sum_{k=0}^N |k,N-k\rangle,
\end{equation}
and the so-called NOON states 
\begin{equation}
|\Psi\rangle = \frac{1}{\sqrt{2}} \left( |N_a, 0\rangle + |0, N_b\rangle \right).
\end{equation}
States with the latter form can be used to beat the standard quantum limit of measurements of phase (or frequency) \cite{Dowling08}, and both represent generalizations of the Bell states of two qubits or the highly nonclassical N-particle Greenberger-Horne-Zeilinger state. 

\begin{figure}[b]
\begin{center}
\includegraphics[width=3in]{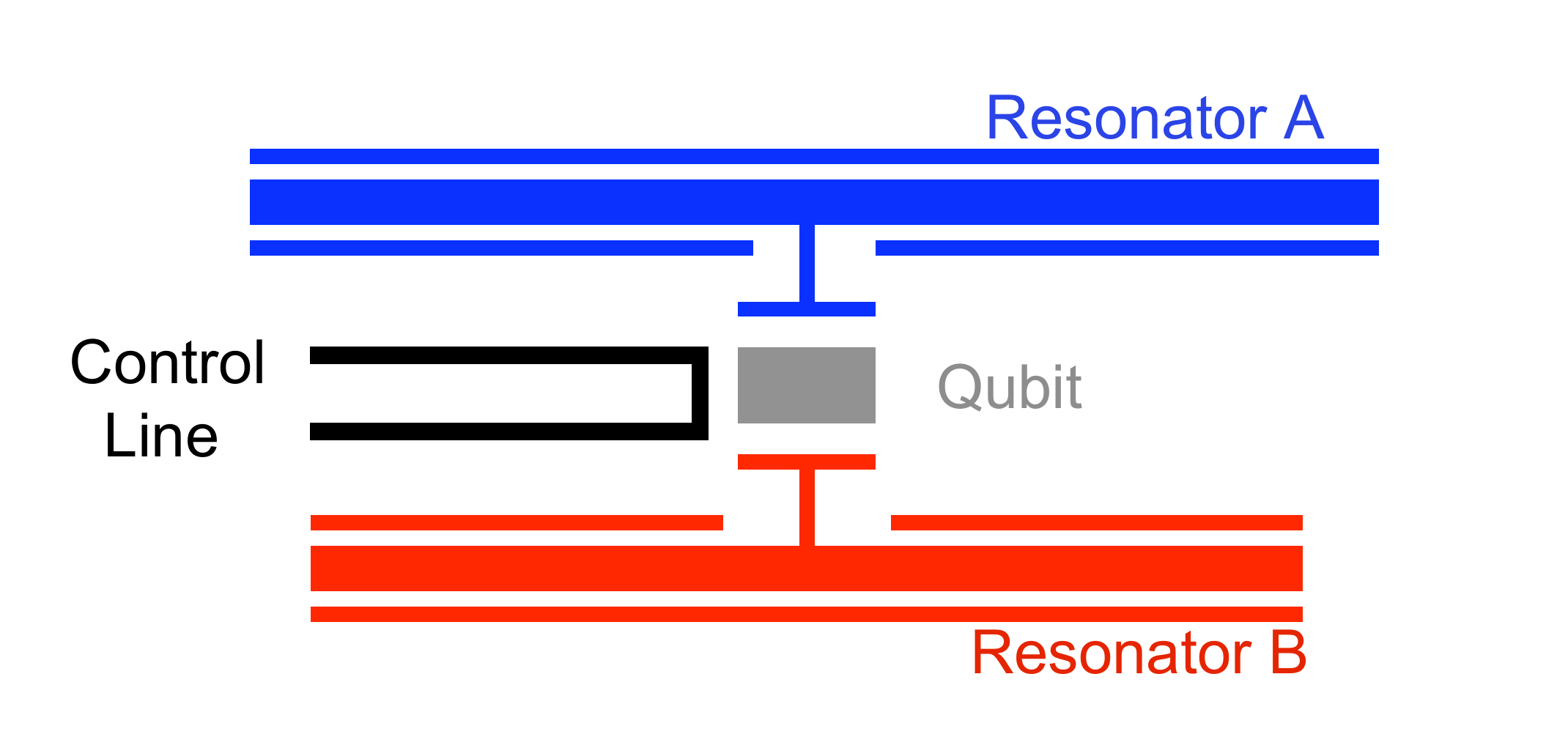}
\caption{Schematic circuit for generating entanglement between two superconducting resonators.  Resonator A (blue) has a fundamental frequency $\omega_a/2\pi$, while resonator B (red) has frequency $\omega_b/2\pi$.  These are each capacitively coupled to a tunable qubit (gray) with frequency $\omega_q/2\pi$, with coupling strengths $g_a$ and $g_b$.  The qubit is controlled by an external circuit (black).  The theoretical results described in the text require $\omega_a < \omega_q < \omega_b$.}
\label{circuitfig}
\end{center}
\end{figure}

In this Letter we present a realistic solution to the state-synthesis problem for two superconducting resonators.  We consider a tunable superconducting qubit, such as the phase \cite{Martinis2002} or transmon \cite{Koch07} qubit, coupled to two resonators with different frequencies, as shown in Fig.~\ref{circuitfig}.  This is described by the Hamiltonian 
\begin{equation}
\begin{array}{lcl}
H &=& \omega_q(t) |1\rangle \langle 1| +  \frac{1}{2} \left( \Omega(t) |1\rangle \langle 0| + \Omega^*(t) |0\rangle \langle 1| \right)   \\
& & +  \omega_a a^{\dagger} a + \omega_b b^{\dagger} b \\
& & + g_a \left( \sigma_+ a  + \sigma_- a^{\dagger} \right) + g_b \left( \sigma_+ b  + \sigma_- b^{\dagger} \right),
\end{array} 
\label{rqr}
\end{equation}
where $a^{\dagger}$ is the creation operator for the resonator of frequency $\omega_a$, $b^{\dagger}$ is the creation operator for a resonator of frequency $\omega_b$, $\Omega(t)$ is a possibly complex microwave field (in the rotating wave approximation), and $g_a$ and $g_b$ are fixed coupling strengths between the qubit and the resonators.  Control of this circuit is exercised by modifying the time dependent qubit frequency $\omega_q(t)$ by a ``shift'' pulse and applying Rabi pulses resonant with the qubit.  By performing a sequence of these pulses, quanta can be created and transferred between the qubit and the two resonators.  

Note that this resonator-qubit-resonator system complements the qubit-resonator-qubit systems first studied in the resonant regime ($\omega_q = \omega_a$) at NIST~\cite{Sillanpaa2007} and the dispersive regime ($|\omega_q - \omega_a| \gg g_a$) at Yale~\cite{Majer07} (note also the recent experiments \cite{Leek2010, Johnson2010}).   There have been several theoretical studies of this type of system \cite{Xue07b}, all towards the goal of generating entanglement between mesoscopic resonators.  Here we solve the general problem of  synthesizing an arbitrary entangled state, an important step towards quantum information processing with harmonic oscillator modes. 

A state-synthesis algorithm must program a sequence of pulses to prepare and transfer Fock states into the desired superposition state.  Previous studies of this problem for entangled states of motion for a single trapped ion~\cite{Gardiner97} have shown that the synthesis of a general state with $n_a \le N_{\mbox{\scriptsize max}}$ and $n_b \le N_{\mbox{\scriptsize max}}$ requires a number of elementary steps of order $N_{\mbox{\scriptsize max}}^2$, proportional to the number of coefficients in the expansion of the state vector, and schemes that achieve this scaling have been identified.  However, none of these schemes can be directly applied to the problem presented above.  These schemes all use sideband transitions and most use special two-mode interactions~\cite{Steinbach97} specific to ion traps.  While there are sideband interactions for resonators dispersively coupled to a qubit~\cite{Blais2007}, these interactions will be much slower; the transfer of a single photon has an effective Rabi coupling of $\Omega_{\mbox{\scriptsize eff}} \sim g |\Omega|^2/ \omega_q^2$~\cite{Wallraff2007}.  

\begin{figure*}
\begin{center}
\includegraphics[width=6in]{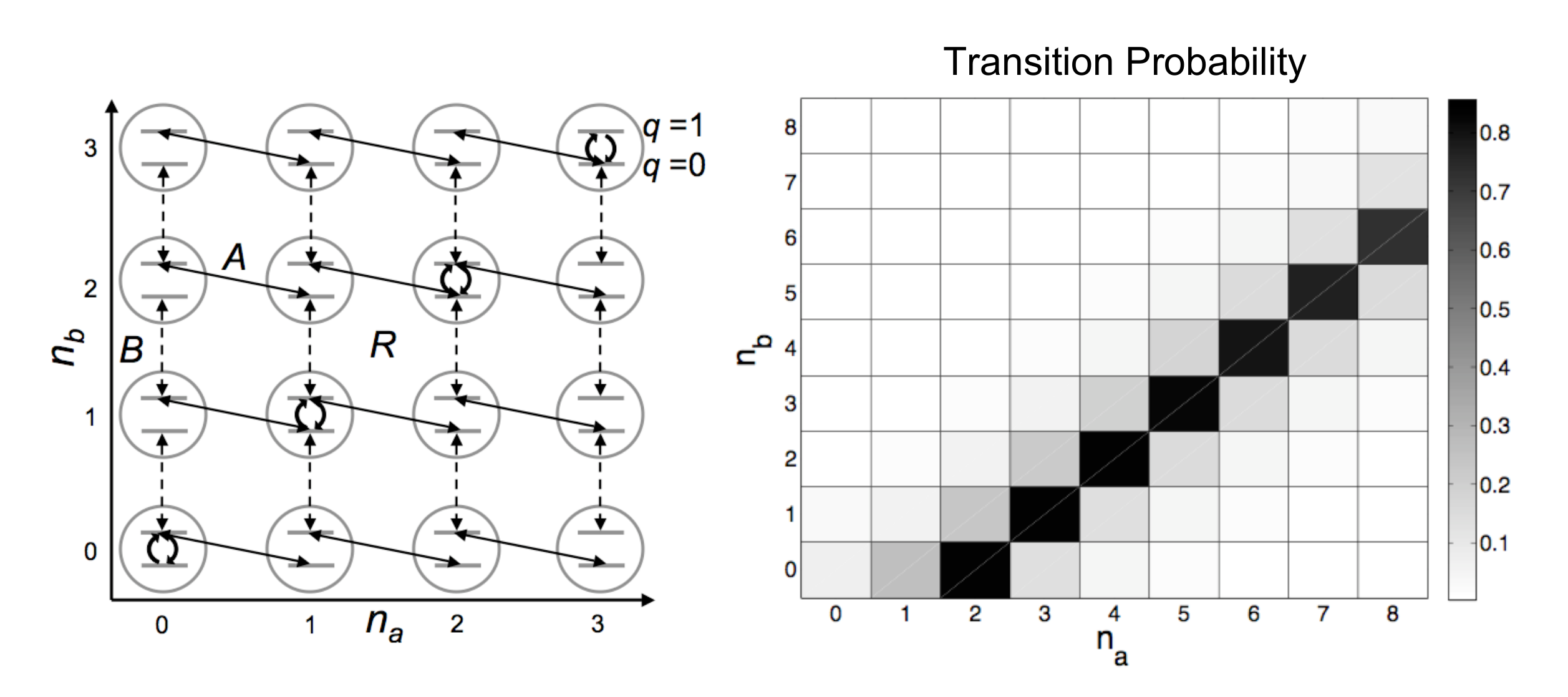}
\caption{Schematic set of operations to generate an arbitrary state of two resonators.  In this Fock-state diagram, the state $|q,n_a,n_b\rangle$ is represented by the node at location $(n_a,n_b)$.  (a) Interactions lead to couplings between these states, indicated by the arrows.  Three key interactions are used:  $A$ transfers quanta between the qubit and resonator $a$ (solid lines),  $B$ transfers quanta between the qubit and resonator $b$ (dashed lines), and $R$ (curved arrows) rotates the qubit for states with $n_a - n_b = n$ (here with $n=0$, see text).   (b) Numerical simulations of Stark-shifted Rabi oscillations for $\omega_a / (2\pi) = 6.3 \ \mbox{GHz}$, $\omega_b/(2\pi) = 7.7\ \mbox{GHz}$, $\omega_q /(2\pi) = 7\ \mbox{GHz}$, $g_a/(2\pi) = g_b/(2\pi) =  70  \ \mbox{MHz}$.  The Rabi oscillations are driven at $\omega_d/ (2\pi) = 7.025 \ \mbox{GHz}$ and $\Omega/(2\pi) = 7 \ \mbox{MHz}$ (with $n_a - n_b = 2$). Each block corresponds to the maximum probability of the transition $|0,n_a, n_b\rangle \to |1, n_a,n_b\rangle$ (see text).}
\label{Fock_Diagram}
\end{center}
\end{figure*}

For definiteness, consider the Fock-state diagram of Fig.~\ref{Fock_Diagram}(a).  Each node represents the two states of the qubit with photon numbers $(n_a,n_b)$, or the quantum state $|q,n_a,n_b\rangle$ (where the qubit state is $q=0$ or $1$).  The resonant interaction of the qubit with each resonator, $H_a = g_a (\sigma_+ a + \sigma_- a^{\dagger})$ and $H_b = g_b (\sigma_+ b + \sigma_- b^{\dagger})$, leads to horizontal and vertical transitions, respectively, illustrated by the solid and dashed lines.  These resonant interactions are fast, efficient, and provide nearest neighbor transitions in the Fock-state diagram.  The addition and control of individual photons requires an independent state-selective qubit rotation.  By addressing the qubit in between the resonant cavity interactions, it is possible to access the entire state space.

To achieve selective manipulations of the quantum system, we use the photon-number-dependent Stark shift \cite{Blais04}.  For our system, this implies that a qubit operated in the dispersive regime will undergo Rabi oscillations from  $|0,n_a,n_b\rangle \to |1, n_a,n_b\rangle$ when the drive frequency satisfies
\begin{equation}
\omega_d = \omega_q + \frac{g_a^2}{\omega_q - \omega_a} (2 n_a + 1) +   \frac{g_b^2}{\omega_q - \omega_b} (2 n_b +1).
\label{dispersive}
\end{equation}
We choose to set $\Delta \omega  = 2g_a^2/(\omega_q-\omega_a) = -2g_b^2/(\omega_q- \omega_b)$.  This can always be achieved for a qubit with a tunable frequency such as the phase or transmon qubit.  This allows us to simplify our Rabi pulses to frequencies
\begin{equation}
\omega_n = \omega_q + n \Delta \omega,
\end{equation}
which selects those states with $n_a - n_b = n$, where $n$ is an integer ($n=0$ is shown in Fig.~\ref{Fock_Diagram} (a)).  Note that this choice requires $\omega_a < \omega_q < \omega_b$ and $|\Omega|<g_a^2/(\omega_q-\omega_b)$ (to avoid nonresonant transitions).   By choosing different values of $\omega_n$ one can address each of the ``diagonals'' of the Fock-state diagram.  

Direct simulations of Rabi oscillations using the full Hamiltonian verify this approach, as shown in Fig.~\ref{Fock_Diagram}(b) for $n=2$.  Each block represents the maximum of the transition probabilities $ |\langle 1, n_a,n_b | \Psi(t)\rangle|^2$, calculated with the initial condition $|\Psi(t=0)\rangle = |0, n_a, n_b \rangle$.  The transition probabilities are large along the diagonal, decreasing significantly for neighboring Fock states.  A similar effect known as ``number splitting'' was experimentally demonstrated for a qubit coupled to a single resonator \cite{Gambetta06}, and recently used in an experiment with a qubit coupled to two resonators \cite{Johnson2010}.  

We now show how these three interactions can be used to achieve the synthesis of arbitrary joint states of two resonators.  This is accomplished by the following sequence of operations 
\begin{equation}
 U = \left(\prod_{j=1}^{N_b} U_{b,j} \right) U_{a}
 \end{equation}
 with
\begin{equation}
U_{a} =  \prod_{j=1}^{N_a} A_j R_{a,j}  \  \mbox{and} \ \ U_{b,j} =  \prod_{k=0}^{N_a} B_{jk} R_{b,jk}.
\end{equation}
Here we have defined $A_j = \exp(-i H_a t_{a,j})$, $B_{j} = \exp(-i H_b t_{b,jk})$, and the single-qubit rotations $R_{a,j}$ and $R_{b,jk}$.  This sequence can be physically realized by shifting the qubit into and out of resonance with resonators $a$ and $b$, interleaved by the Stark-shifted qubit rotations described above.  The parameters ($t_{a,j}$, $t_{b,jk}$, $R_{a,j}$, and $R_{b,jk}$) in this operation must be chosen to satisfy
 \begin{equation}
| \Psi \rangle = U |0, 0,0\rangle = |0\rangle \otimes \sum_{n_a=1}^{N_a} \sum_{n_b=1}^{N_b}  c_{n_a,n_b} |n_a,n_b\rangle,
 \end{equation}
where $c_{n_a,n_b}$ are arbitrary coefficients.  To determine the precise sequence of operations for a given two-resonator state $|\Psi\rangle$, one solves for the inverse evolution: 
\begin{equation}
U^{\dagger} |\Psi\rangle = U_{a}^{\dagger} \prod_{j=N_b}^1 U_{b,j}^{\dagger} |\Psi\rangle = |0,0,0\rangle.
\label{inverse}
\end{equation} 

Each step of this inverse sequence can be mapped onto a state transfer in the Fock-state diagram.  In order to solve this problem, one must show that photons can be consistently removed from the system.  Our approach accomplishes this in the following way.  The sequence of $U_{b,j}^{\dagger}$ unitaries moves the system along the vertical paths ($B$) in the Fock-state diagram.  Each step removes a photon from row $n_b = j$ of the Fock-state diagram.  This is done by marching (from right to left) along the columns $n_a = k$, with $B_{jk}$ transferring the amplitude in row $j$ to row $j-1$, after which the Stark-shifted single-qubit operations  $R_{b,jk}$ (with frequencies $\omega_q + (k-j +1) \Delta \omega$) rotates the amplitude to state $|0,k,j-1\rangle$.  After all of the photons have been removed from the columns in row $j$, the sequence repeats for row $j-1$.  Each time through, population is transferred towards $n_b = 0$.  Once there, the $U_a^{\dagger}$ sequence moves population along the horizontal paths ($A$) to $n_a = n_b = 0$, or $|0,0,0\rangle$, thus solving Eq. (\ref{inverse}). This completes the algorithm.

The total number of steps matches the optimal efficiency of the ion-trap proposals discussed above, but here using resonant interactions and the Stark-shifted Rabi pulses.  Each step involves the rotation of an effective two-state system whose amplitudes are known (from the original $c_{n_a,n_b}$), as in the original Law-Eberly scheme \cite{Law96}.  By counting the number of operations in $U$, we find that the general sequence requires $N_a$ $A$ unitaries, $(N_a+1) N_b$ $B$ unitaries, and $N_a + (N_a + 1) N_b$ Rabi pulses.  Thus, the total time is approximately given by
\begin{equation}
T_{\mbox{\scriptsize{max}}} = (N_a +1) (N_b+1) \frac{\pi}{\Omega} + \sum_{j=1}^{N_a} \frac{\pi}{2 g_a \sqrt{j}} + (N_a+1) \sum_{j=1}^{N_b} \frac{\pi}{2 g_b \sqrt{j}}.
\label{tmax}
\end{equation}

\begin{figure}[t]
\begin{center}
\includegraphics[width=3in]{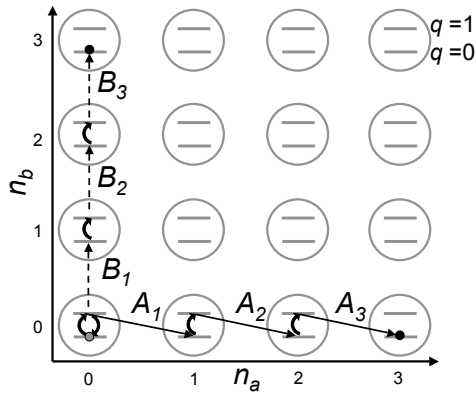}
\caption{Algorithm to generate the state $|\Psi \rangle = |3,0\rangle + |0,3\rangle$ of two coupled resonators.  The sequence of operations is detailed in Table 1.  The horizontal, vertical, and curved transitions are the interactions $A$, $B$, and $R$ (see text).}
\label{noonstatefig}
\end{center}
\end{figure}

Note, however, that for states such as the NOON state we can achieve an even greater efficiency.  For these states one need not transfer amplitude over the whole diagram, but only along certain paths, leading to a sequence with only a linear number of steps.  Consider the sequence of operations shown in Fig.~\ref{noonstatefig}, whose steps are detailed in Table \ref{fockprogram}.  This sequence requires a linear number of operations as opposed to the quadratic scaling of the general procedure described above.  In fact, we find that  Eq.~(\ref{tmax}) can be reduced to
\begin{equation}
T_{\mbox{\scriptsize{NOON}}} = \left(N_a + N_b - \frac{1}{2}\right) \frac{\pi}{\Omega} + \sum_{j=1}^{N_a} \frac{\pi}{2 g_a \sqrt{j}} + \sum_{j=1}^{N_b} \frac{\pi}{2 g_b \sqrt{j}}.
\label{nooneq}
\end{equation}

We now estimate the time required to generate a NOON state.  We consider a qubit with $\omega_q/(2\pi) = 6.5 \ \mbox{GHz}$ and resonators with $\omega_a/(2\pi) = 6 \ \mbox{GHz}$, $\omega_b/(2\pi) = 7 \ \mbox{GHz}$ and a coupling of $g_a/(2\pi) = g_b/(2\pi) = 150 \ \mbox{MHz}$ (similar to recent experiments \cite{DiCarlo2009}).  For the state-selective Rabi pulses, numerical simulations (not shown) show that $\Omega/(2\pi) = \frac{1}{4 \pi} g_a^2/(\omega_q-\omega_a) = 22 \ \mbox{MHz}$ produces an error of a few percent.  Using Eq.~(\ref{nooneq}) we estimate that NOON state generation with $N_a = N_b = 8$ will take only $360 \ \mbox{ns}$.  Smaller couplings (as in Fig. 2) can still achieve $N_a = N_b = 3$ in $410 \ \mbox{ns}$.  These times compare quite favorably to the coherence times of both qubits and resonators, which are now consistently greater than $0.5 \ \mu\mbox{s}$ \cite{DiCarlo2009, Hofheinz2008, Hofheinz2009}.   Most of the time is for the high-fidelity state-selective Rabi pulses.  Faster rotations should be possible by using specially shaped pulses \cite{Steffen2003}.  Recent experiments \cite{Chow09} using such pulses show that quantum algorithms are ultimately limited by the coherence times.

Other experimental issues may arise in this procedure.  First, there will be modifications to the rotating wave and dispersive approximations (used to derive Eq.(\ref{dispersive})) for large couplings and photon numbers.  These can be addressed through pulse shaping or optimal control theory approaches.  Second, we have ignored the dynamical phases that arise when the qubit is shifted between frequencies.  These phases can be corrected by including brief pauses between the Rabi and shift pulses \cite{DiCarlo2009,Hofheinz2009}, and will not significantly affect the overall time needed for state preparation.   A full simulation including these effects, including decoherence, will be performed elsewhere, but the estimates given above are quite promising.   Finally, verifying the two-resonator state may require additional qubits for readout using Wigner tomography \cite{Hofheinz2009} to probe the coherence of the two resonators.  However, efficiently manipulating and measuring entangled resonators appears experimentally possible.

\begin{table}[t] 
\caption{NOON State-synthesis procedure} 
\centering      
\begin{tabular}{l l r}  
\hline\hline                        
Step & Parameters & Quantum State \\ [0.5ex] 
\hline                    
$R_{a,1} $ & $\Omega t_{qa,1}= \pi/2, \omega_d = \omega_0$ & $|0,0,0\rangle - i |1,0,0\rangle$  \\ 
$A_{1} $ & $g_a t_{a,1}= \pi/2$ & $|0,0,0\rangle -  |0,1,0\rangle$  \\ 
$R_{a,2} $ & $\Omega t_{qa,2}= \pi , \omega_d = \omega_1$ & $|0,0,0\rangle + i |1,1,0\rangle$  \\ 
$A_{2} $ & $g_a t_{a,2}= \pi/(2\sqrt{2})$ & $|0,0,0\rangle +  |0,2,0\rangle$  \\ 
$R_{a,3} $ & $\Omega t_{qa,3}= \pi , \omega_d = \omega_2$ & $|0,0,0\rangle - i |1,2,0\rangle$  \\ 
$A_{3} $ & $g_a t_{a,3}= \pi/(2\sqrt{3})$ & $|0,0,0\rangle -  |0,3,0\rangle$  \\ 
$R_{b,1} $ & $\Omega t_{qb,1}= \pi, \omega_d = \omega_0$ & $-i |1,0,0\rangle - |0,3,0\rangle$  \\ 
$B_{1} $ & $g_b t_{b,1}= \pi/2$ &  $-|0,0,1\rangle -  |0,3,0\rangle$  \\ 
$R_{b,2} $ & $\Omega t_{qb,2}= \pi, \omega_d = \omega_{-1}$ & $i |1,0,1\rangle - |0,3,0\rangle$  \\ 
$B_{2} $ & $g_b t_{b,2}= \pi/(2\sqrt{2})$ & $|0,0,2\rangle -  |0,3,0\rangle$  \\ 
$R_{b,3} $ & $\Omega t_{qb,3}= \pi, \omega_d = \omega_{-2}$ & $-i |1,0,2\rangle - |0,3,0\rangle$  \\ 
$B_{3} $ & $g_b t_{b,3}= \pi/(2\sqrt{3})$ & $-|0,0,3\rangle -  |0,3,0\rangle$  \\ 
[1ex]       
\hline     
\end{tabular} 
\label{fockprogram}  
\end{table} 

In summary, we have presented a method to synthesize an arbitrary quantum state of two superconducting resonators.   This method combines state-selective Rabi oscillations (using Stark shifts) with linear couplings of each resonator to a tunable artificial atom.  We have shown that this approach can efficiently synthesize NOON states, with large photon numbers, using existing technology.  More generally, this approach can be applied to many types of coupled qubit-resonator systems, and opens up an important path towards quantum information processing with superconducting oscillators.

\acknowledgments
We gratefully acknowledge discussions with J. Aumentado and F. Altomare.  FWS was supported by the Research Corporation for Science Advancement, and KJ by the NSF under Project No.\ PHY-0902906.

\bibliography{report}

\end{document}